\pdfoutput=1

\documentclass[11pt,twoside,a4paper,cmspaper,final,collab]{cms-tdr}
\def\svnVersion{218722}\def\cmsCernNo{2013-057}\def\cmsCernDate{\today}\def\cmsMessage{Published in the European Physical Journal C as \href{http://dx.doi.org/10.1140/epjc/s10052-013-2604-6}{\doi{10.1140/epjc/s10052-013-2604-6.}}}

\begin{document}\cmsNoteHeader{QCD-11-003}

\hyphenation{had-ron-i-za-tion}
\hyphenation{cal-or-i-me-ter}
\hyphenation{de-vices}

\RCS$Revision: 205962 $
\RCS$HeadURL: svn+ssh://svn.cern.ch/reps/tdr2/papers/QCD-11-003/trunk/QCD-11-003.tex $
\RCS$Id: QCD-11-003.tex 205962 2013-09-06 11:28:37Z alverson $
\newlength\cmsFigWidth
\ifthenelse{\boolean{cms@external}}{\setlength\cmsFigWidth{0.49\textwidth}}{\setlength\cmsFigWidth{0.9\linewidth}}
\ifthenelse{\boolean{cms@external}}{\providecommand{\cmsLeft}{top}}{\providecommand{\cmsLeft}{left}}
\ifthenelse{\boolean{cms@external}}{\providecommand{\cmsRight}{bottom}}{\providecommand{\cmsRight}{right}}
\providecommand{\alps}{\ensuremath{\alpha_S}\xspace}
\providecommand{\alpsmz}{\ensuremath{\alpha_S(M_\cPZ)}\xspace}
\providecommand{\alpsq}{\ensuremath{\alpha_S(Q)}\xspace}
\providecommand{\HERWIGPP} {{\textsc{herwig++}}\xspace}
\providecommand{\NLOJETPP} {{\textsc{NLOJet++}}\xspace}
\providecommand{\fastNLO} {{\textsc{fastNLO}}\xspace}
\providecommand{\fastjet} {{\textsc{FastJet}}\xspace}
\providecommand{\PYTHIAS} {{\textsc{pythia6}}\xspace}
\providecommand{\PYTHIAE} {{\textsc{pythia8}}\xspace}
\providecommand{\RooUnfold} {{\textsc{RooUnfold}}\xspace}
\providecommand{\avept}{\ensuremath{\langle p_\mathrm{T1,2}\rangle}\xspace}
\providecommand{\Ratio}{$R_\mathrm{32}$\xspace}

\providecommand{\rbthm}{\rule[-2ex]{0ex}{5ex}}
\providecommand{\rbtrr}{\rule[-0.8ex]{0ex}{3.2ex}}

\cmsNoteHeader{QCD-11-003} 
\title{Measurement of the ratio of the inclusive 3-jet cross section to the inclusive 2-jet cross section in pp collisions at $\sqrt{s} = 7\TeV$ and first determination of the strong coupling constant in the \TeV range
}
\titlerunning{Measurement of the ratio of the inclusive 3-jet cross section to the inclusive 2-jet cross section}

\date{\today}

\abstract{
  A measurement is presented of the ratio of the inclusive 3-jet cross
  section to the inclusive 2-jet cross section as a function of the
  average transverse momentum, $\langle p_\mathrm{T1,2}\rangle$, of
  the two leading jets in the event. The data sample was collected
  during 2011 at a proton-proton centre-of-mass energy of
  $7\TeV$ with the CMS detector at the LHC, corresponding to
  an integrated luminosity of $5.0\fbinv$. The strong
  coupling constant at the scale of the \cPZ\ boson mass is determined
  to be ${\alpha_{S}(M_\cPZ) = 0.1148 \pm 0.0014\,\text{(exp.)} \pm
  0.0018\,\text{(PDF)} \pm 0.0050\,\text{(theory)}}$, by
  comparing the ratio in the range $0.42 < \langle
  p_\mathrm{T1,2}\rangle < 1.39\TeV$ to the predictions of
  perturbative QCD at next-to-leading order. This is the first
  determination of $\alpha_{S}(M_{\cPZ})$ from measurements at
  momentum scales beyond $0.6\TeV$. The predicted ratio
  depends only indirectly on the evolution of the parton distribution
  functions of the proton such that this measurement also serves as a
  test of the evolution of the strong coupling constant.
  No deviation from the expected behaviour is observed.
}

\hypersetup{%
  pdfauthor={CMS Collaboration},%
  pdftitle={Measurement of the ratio of the inclusive 3-jet cross
    section to the inclusive 2-jet cross section in pp collisions at
    sqrt(s) = 7 TeV and first determination of the strong coupling
    constant in the TeV range},%
  pdfsubject={CMS},%
  pdfkeywords={CMS, physics, QCD, multijets ratio, strong coupling constant}%
}

\maketitle 

\section{Introduction\label{section:intro}}

As a consequence of the non-Abelian nature of quantum chromodynamics
(QCD), the renormalization group equation
(RGE)~\cite{bib:Callan,bib:Symanzik1,bib:Symanzik2} predicts that the
strong force becomes weaker at short distances corresponding to large
momentum transfers, a property of QCD referred to as asymptotic
freedom. The strength of the strong force, \alpsq, at a given distance
or momentum scale $Q$ is not predicted and has to be extracted from
experiment. Measurements at different $Q$ can then be compared for
consistency with QCD via the RGE, which precisely describes the
evolution of $\alpha_S(\mu_r)$, where the renormalization scale
$\mu_r$ is identified with $Q$.  By convention, the consistency is
tested by evolving all values of \alpsq to the common scale ${\mu_r =
  Q = M_\cPZ}$, i.e.\ the precisely known mass of the $\cPZ$
boson. The current world average value is ${\alpsmz = 0.1184 \pm
  0.0007}$~\cite{bib:PDG2012}.

Measurements of the \textit{running}\/ of \alpsq provide a stringent test
of QCD\@.  Previous collider experiments at LEP and HERA have
established the validity of the RGE up to momentum transfers $Q$ of
$208\GeV$~\cite{bib:PDG2012}. A recent publication by the D0
Collaboration extends this range up to
$400\GeV$~\cite{bib:as_D0_angular}. The determination of \alpsq from
jet cross sections as in~\cite{bib:as_CDF}
or~\cite{bib:as_ATLAS_xsection} depends directly on parton
distribution functions (PDFs) that have been evolved from small to
very high momentum scales via the
Dokshitzer--Gribov--Lipatov--Altarelli--Parisi (DGLAP)
equations~\cite{bib:Gribov,bib:Altarelli,bib:Dokshitzer}, which assume
the validity of the RGE\@. This dependence on the evolution of the
PDFs can be reduced by investigating cross section ratios. The ratio
\Ratio of the inclusive 3-jet cross section to the inclusive 2-jet
cross section is proportional to \alpsq where $Q$ is defined as the
average transverse momentum of the two jets leading in $\pt$,

\begin{equation}
  Q = \avept = \frac {p_\mathrm{T1}+p_\mathrm{T2}}{2}.
\end{equation}

Many theoretical systematic uncertainties related to the choice of the
renormalization and factorization scales, $\mu_r$ and $\mu_f$, or to
nonperturbative effects are reduced in the cross section ratio. In
addition, experimental uncertainties such as those due to the jet
energy scale largely cancel in the measurement of \Ratio. The
uncertainty on the integrated luminosity measurement cancels
completely.  The Compact Muon Solenoid (CMS) Collaboration has
previously measured \Ratio~\cite{bib:R32_2011}, and the predictions of
various Monte Carlo (MC) event generators were found to be in general
agreement with the measurement.

This measurement is performed using a sample of multijet events,
collected during 2011 by the CMS experiment at the Large Hadron
Collider (LHC), corresponding to an integrated luminosity of
$5.0\fbinv$ of pp collisions at a
centre-of-mass energy $\sqrt{s}=7\TeV$. The transverse momentum $\pt$
and the rapidity $y$ of a jet with energy $E$ and momentum
$\vec{p}=(p_x,p_y,p_z)$ (where $p_z$ is the momentum component along
the direction of the anticlockwise proton beam) are defined as $\pt =
\sqrt{\smash[b]{p_x^2+p_y^2}}$ and $y = \frac{1}{2} \ln [(E+p_{z})/ (E-p_{z})] $,
respectively. Jets are reconstructed using the infrared- and
collinear-safe anti-\kt clustering
algorithm~\cite{bib:antikt,bib:fastjet} with a size parameter of 0.7.
This measurement uses jets with $\pt > 150\GeV$ and $\vert y \vert <
2.5$.

The large number of multijet events collected over a wide range of
\avept, $420<\avept<1390\GeV$, allows \alpsq to be determined with
only a small dependence on the evolution of the PDFs, thus testing the
validity of the RGE in an extended range of transverse momenta.

\section{The CMS detector\label{section:detector}}

The central feature of the CMS
apparatus is a superconducting solenoid, 13\unit{m} in length and 6\unit{m} in
diameter, providing an axial magnetic field of 3.8\unit{T}. The field volume
of the solenoid is instrumented with various layers of particle
detection systems. Charged particle trajectories are measured by the
silicon pixel and strip tracker, with full azimuthal coverage within
$\vert \eta \vert < 2.5$, where the pseudorapidity $\eta$ is defined
as $\eta = -\ln[\tan(\theta/2)]$, and $\theta$ is the polar angle
with respect to the $z$ axis. Surrounding the
trackers are a lead tungstate crystal electromagnetic calorimeter
(ECAL) with a preshower detector in the endcaps, and a
brass and scintillator hadron calorimeter (HCAL), covering the region
$\vert \eta \vert < 3$. In addition to the barrel and endcap
detectors, CMS has extensive forward calorimetry which extends the
coverage to $ \vert \eta \vert = 5 $. The steel flux return yoke
outside the solenoid is instrumented with gas-ionization detectors
used to identify and reconstruct muons.
A more detailed description of the CMS detector can be found
in~\cite{bib:CMS_experiment}.

\section{Event selection and
  reconstruction\label{section:reconstruction}}

The CMS detector records events using a two-level trigger system
consisting of a hardware-based level-1 (L1) trigger and a
software-based high level trigger (HLT). In this study, single-jet
triggers that reconstruct jets from calorimeter energy deposits at L1
and HLT are used to select events based on three HLT \pt thresholds,
190, 240, and 370\GeV. All except the highest-threshold trigger were
prescaled during the 2011 run. The corresponding integrated
luminosity \lumi for each of the three samples is shown in
Table~\ref{tbl:TrigLumi}. The efficiency of each of the triggers is
estimated using lower-\pt-threshold triggers.
These three jet trigger thresholds ensure 100\% trigger efficiency
in the three jet samples for $\avept > 215$, 269, and $409\GeV$.

\begin{table}[htbH]
  \begin{center}
    \topcaption{The integrated luminosity for each trigger
      sample.\label{tbl:TrigLumi}}
    \begin{tabular}{ccccc}
      \hline\hline
      HLT \pt threshold (\GeVns{})\rbthm   &  190   &  240   &  370    \\
      \hline
      \lumi ($\text{fb}^{-1}$)\rbtrr &  0.15  &  0.51  &  5.0    \\
      \hline\hline
    \end{tabular}
  \end{center}
\end{table}

Each event is required to have at least one offline-reconstructed
vertex~\cite{bib:vertex} along the beam line that is within 24 cm of
the nominal interaction point. The four-vectors of particle
candidates reconstructed by the CMS global event reconstruction
algorithm (also called particle-flow event
reconstruction~\cite{bib:pf1}) are used as input to the jet-clustering
algorithm. The clustering is performed by the \fastjet
package~\cite{bib:fastjet} using four-momentum summation. The global
event reconstruction algorithm reconstructs and identifies each particle with an
optimized combination of subdetector information. The energy of
photons is obtained directly from the ECAL measurements after being
corrected for zero-suppression effects. The energy of electrons is
determined from a combination of the track momentum at the main
interaction vertex, the corresponding ECAL cluster energy, and the
energy sum of all bremsstrahlung photons attached to the track. The
energy of muons is derived from the corresponding track momentum. The
energy of charged hadrons is determined from a combination of the
track momentum and the corresponding ECAL and HCAL energies, corrected
for zero-suppression effects, and calibrated for the nonlinear
response of the calorimeters. Finally, the energy of neutral hadrons
is obtained from the corresponding calibrated ECAL and HCAL energies.

Jet energy corrections~\cite{bib:cms-jes} are derived using simulated
events, generated by \PYTHIA 6.4.22~\cite{Sjostrand:2006za} and
processed through the CMS detector simulation based on
\GEANTfour~\cite{bib:geant}, and in situ measurements with dijet,
photon+jet, and Z+jet events. An offset correction is applied to take
into account the extra energy clustered into jets from additional
proton-proton interactions within the same or neighbouring bunch
crossings (in-time and out-of-time pileup)~\cite{bib:cms-jes}. Pileup
effects are important only for low-\pt jets and become negligible for
jets with $\pt> 200\GeV$. The current measurement is therefore largely
insensitive to pileup effects. The jet energy corrections, which
depend on the $\eta$ and \pt of the jet, are applied to the
jet four-momentum vector as a multiplicative factor.
The multiplicative factor is in general smaller than 1.2, approximately
uniform in $\eta$, with typical values of 1.1 for jets having $\pt=100\GeV$
and decreasing to 1.0 for higher values of \pt.

To suppress nonphysical jets, i.e.\ jets resulting from noise in the
ECAL and/or HCAL calorimeters, tight identification
criteria are applied: each jet should contain at least two particles,
one of which is a charged hadron, and the jet energy fraction carried
by neutral hadrons, photons, muons, and electrons should be less than
90\%. These criteria have an efficiency greater than 99\% for
genuine jets.

The selection of multijet events requires two or more jets
with transverse momentum greater than 150\GeV and $\vert y \vert <
2.5$.
The final sample is extracted by rejecting events if either or both
of the leading jets in \pt have $\vert y \vert > 2.5$.

\section{Measurement of \texorpdfstring{\Ratio}{R(32)} and comparison
  with theoretical predictions\label{section:measurement}}

The measured ratio \Ratio as a function of \avept is the ratio of the
number of selected inclusive 3-jet events to the number of selected
inclusive 2-jet events in each \avept bin.
The ratio \Ratio is corrected for detector smearing effects and unfolded
to stable-particle level.
The unfolding method is the iterative Bayesian
method~\cite{bib:BayesUnfold}, as implemented in the \RooUnfold
software package~\cite{bib:RooUnfold}. Unfolding uses a response
matrix that maps the true distribution onto the measured one.
The response matrix is derived from a simulation, which uses as input
the true \Ratio distribution from \PYTHIAS tune Z2 and introduces the
smearing effects by taking into account the \avept resolution~\cite{bib:cms-jes}.
After unfolding \Ratio to
stable-particle level, the final statistical uncertainties include the
correlation among the various \avept bins.

Two main sources of systematic uncertainties on \Ratio are considered.
The first is due to the jet energy scale (JES) and the second due
to the unfolding.

The JES uncertainty has been estimated to be 2.0--2.5\% for
particle-flow jets~\cite{bib:JES_2011}, depending on the jet \pt and
$\eta$. All mutually uncorrelated JES uncertainty sources are
considered following the procedure described in
Ref.~\cite{bib:QCD004arXiv}. The total systematic uncertainty on
\Ratio due to the JES uncertainty is 1.2\%.

The unfolding method takes into account three different mutually
uncorrelated uncertainty sources. The first arises from
insufficient knowledge of the simulated inclusive 3-jet and 2-jet
\avept spectra, which are used to construct the simulated ratio \Ratio
used in the unfolding. The uncertainty is estimated by varying the
3-jet and 2-jet spectra slopes by $\pm$10\%. This is a conservative
estimate and is motivated by the observed difference in the 3-jet and
2-jet spectra slopes between simulations using the event generators
\PYTHIAS tune Z2 and
\HERWIGPP~\cite{bib:Herwigpp} with its default tune in version 2.3. Simulations
of 3-jet and 2-jet spectra using the
\MADGRAPH~\cite{bib:Madgraph,bib:Madgraph2} event generator are in
agreement with those of \PYTHIAS\@. The second uncertainty arises from
the insufficient knowledge of the \avept resolution and is estimated
by varying it by $\pm$10\%.
This variation is motivated by the observed
difference between data and simulation in the jet energy
resolution~\cite{bib:cms-jes}.
Finally, the third uncertainty arises
from non-Gaussian components in the \avept resolution
and is estimated by adding non-Gaussian tails to the simulation. The overall
systematic uncertainty on \Ratio due to unfolding is less than 1\%.
A potential bias originating from the unfolding technique
is studied by comparing the unfolding result of the Bayesian
method with that of the singular-value decomposition (SVD)
method~\cite{bib:SVD}. The bias is found to be negligible.

The theoretical calculation of the ratio \Ratio is based on
next-to-leading-order (NLO) perturbative QCD (pQCD) calculations
multiplied by a nonperturbative factor, which corrects for multiparton
interactions (MPI) and hadronization effects. The NLO calculations
assume $N_f=5$ massless quark flavours and are based on the
parton-level generator \NLOJETPP~\cite{bib:NLO1,bib:NLO2}.
The computations with \NLOJETPP are performed within the
\fastNLO framework \cite{bib:NLO3,bib:NLO4} using the following four
PDF sets: NNPDF2.1 \cite{bib:NNPDF2.1,bib:NNPDF_Juan},
ABM11 \cite{bib:ABM11}, MSTW2008
\cite{bib:MSTW2008NLO,bib:MSTW2008as}, and
CT10 \cite{bib:CT10,bib:CT10N}. In each case the PDF version employing
next-to-next-to-leading order (NNLO) evolution code is chosen, and for
comparisons the respective default values of \alpsmz, which are 0.119,
0.1134, 0.1171, and 0.118, are used. NNPDF2.1, MSTW2008, and CT10
utilize a variable-flavour-number scheme with the maximal number of
flavours $N_{f,\text{max}}$ equal to 6, 5, and 5 respectively, while
the ABM11 PDF set was developed in a fixed-flavour-number scheme
with $N_f=5$. The renormalization and factorization scales are set to
the average transverse momentum \avept.

The nonperturbative effects are estimated using the \PYTHIAS tune Z2
and \HERWIGPP tune 2.3 event generators. The chosen MC models feature
different descriptions of the phenomena and are representative of the
possible values of the nonperturbative corrections.  The
nonperturbative correction (NPC) factor is defined as the ratio of
\Ratio predicted with the nominal generator settings to that obtained
with the MPI and hadronization switched off. This factor is calculated
considering the average of the two MC generators and has typical
values of $\approx$1.02 for $\avept = 250\GeV$, decreasing to 1.0 for
higher \avept. The uncertainty is considered to be half of the
difference between the NPC values obtained using the two MC generators
and amounts to $\approx$0.1\%, leading to a negligible influence on
the final result.

Finally, uncertainties due to the renormalization and factorization
scales are evaluated by varying from the default choice of
$\mu_{r}=\mu_{f}=\avept$ between $\avept/2$ and $2\avept$,
simultaneously in the numerator and denominator of the ratio \Ratio,
in the following six combinations: $(\mu_{r}/\avept,\mu_{f}/\avept) =
(1/2,1/2)$, $(1/2,1)$, $(1,1/2)$, $(1,2)$, $(2,1)$, and $(2,2)$.

\begin{figure*}[hbtp]
  \begin{center}
        \includegraphics[width=0.47\linewidth]{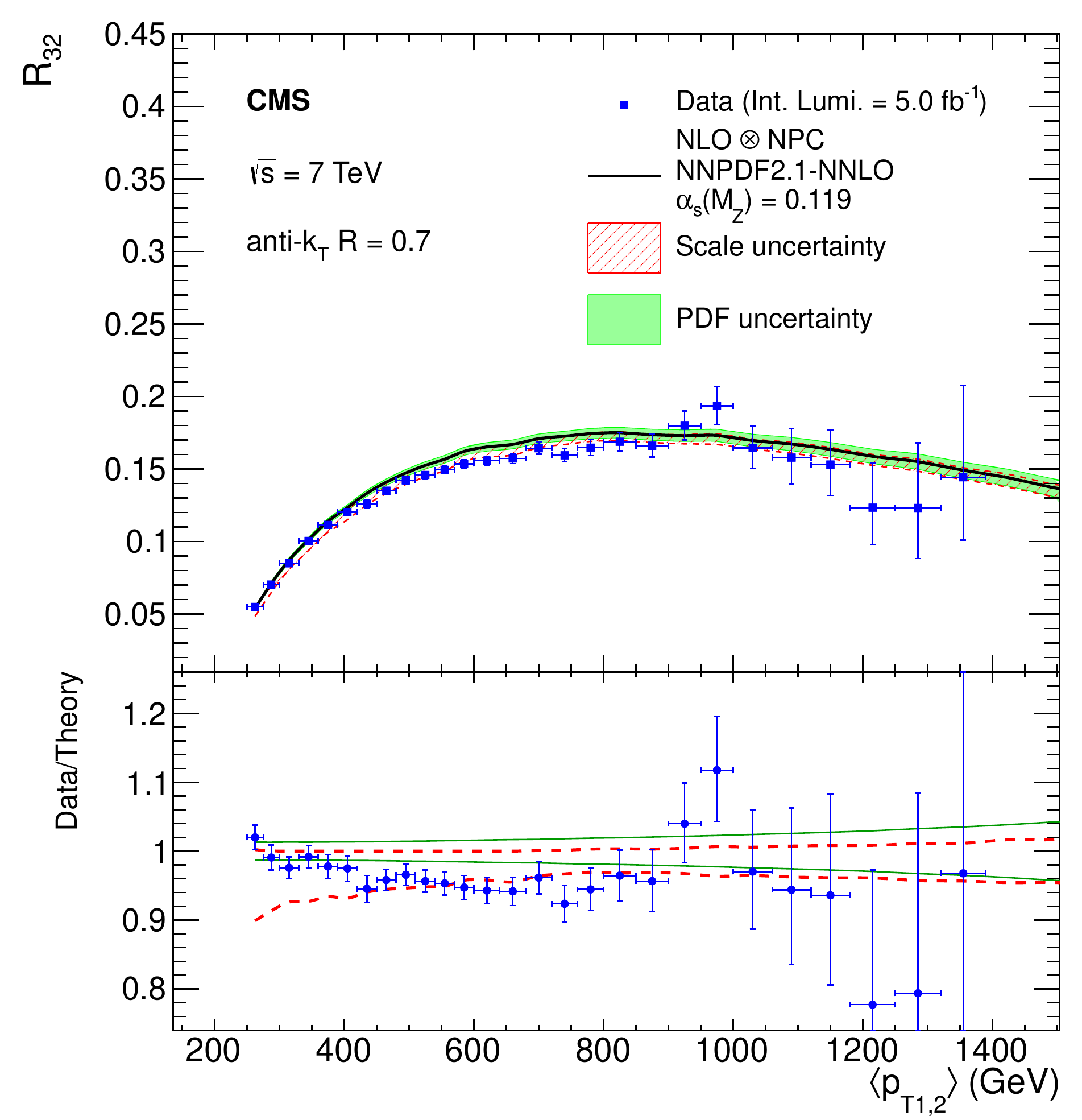}
        \includegraphics[width=0.47\linewidth]{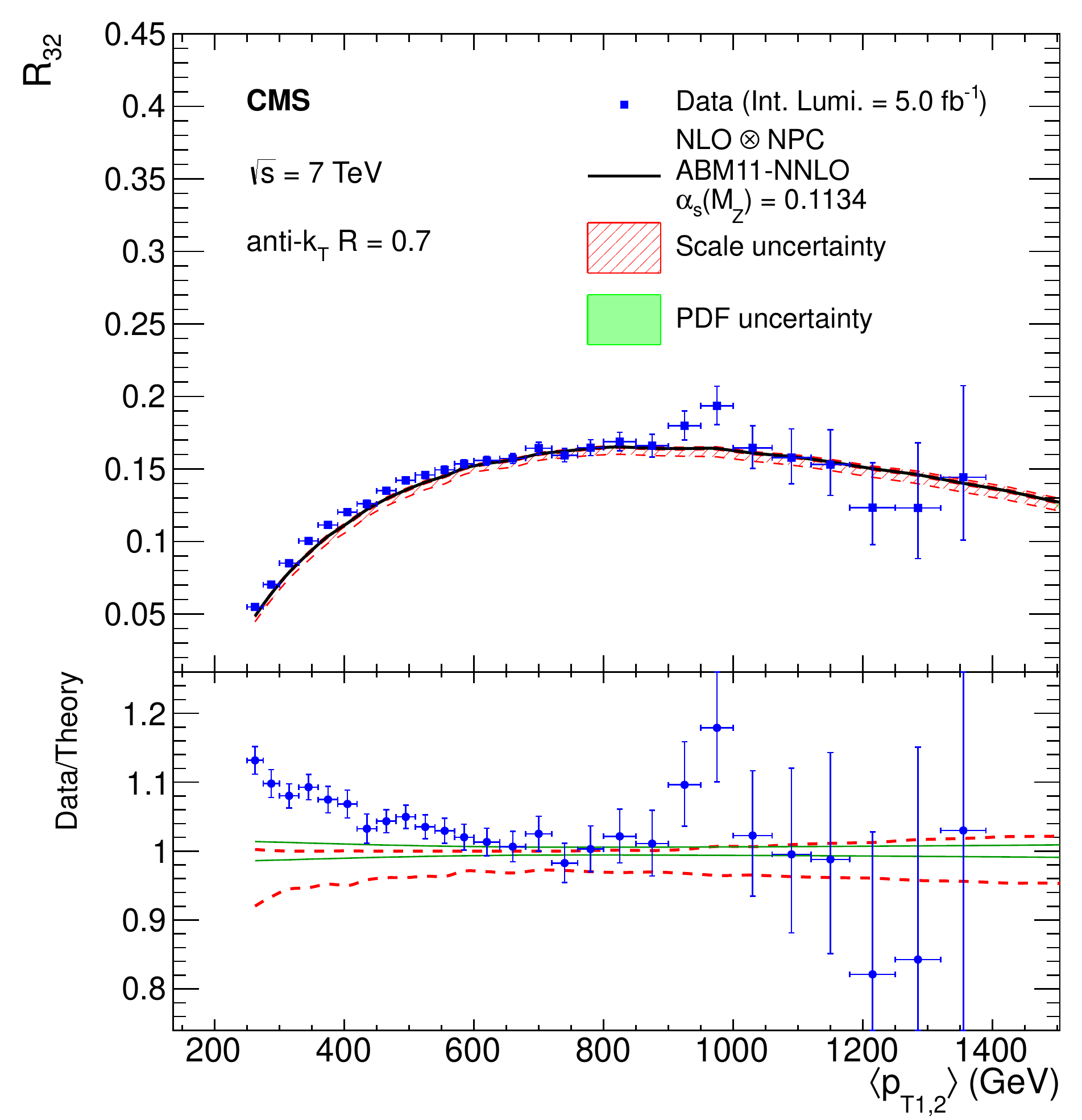}
        \includegraphics[width=0.47\linewidth]{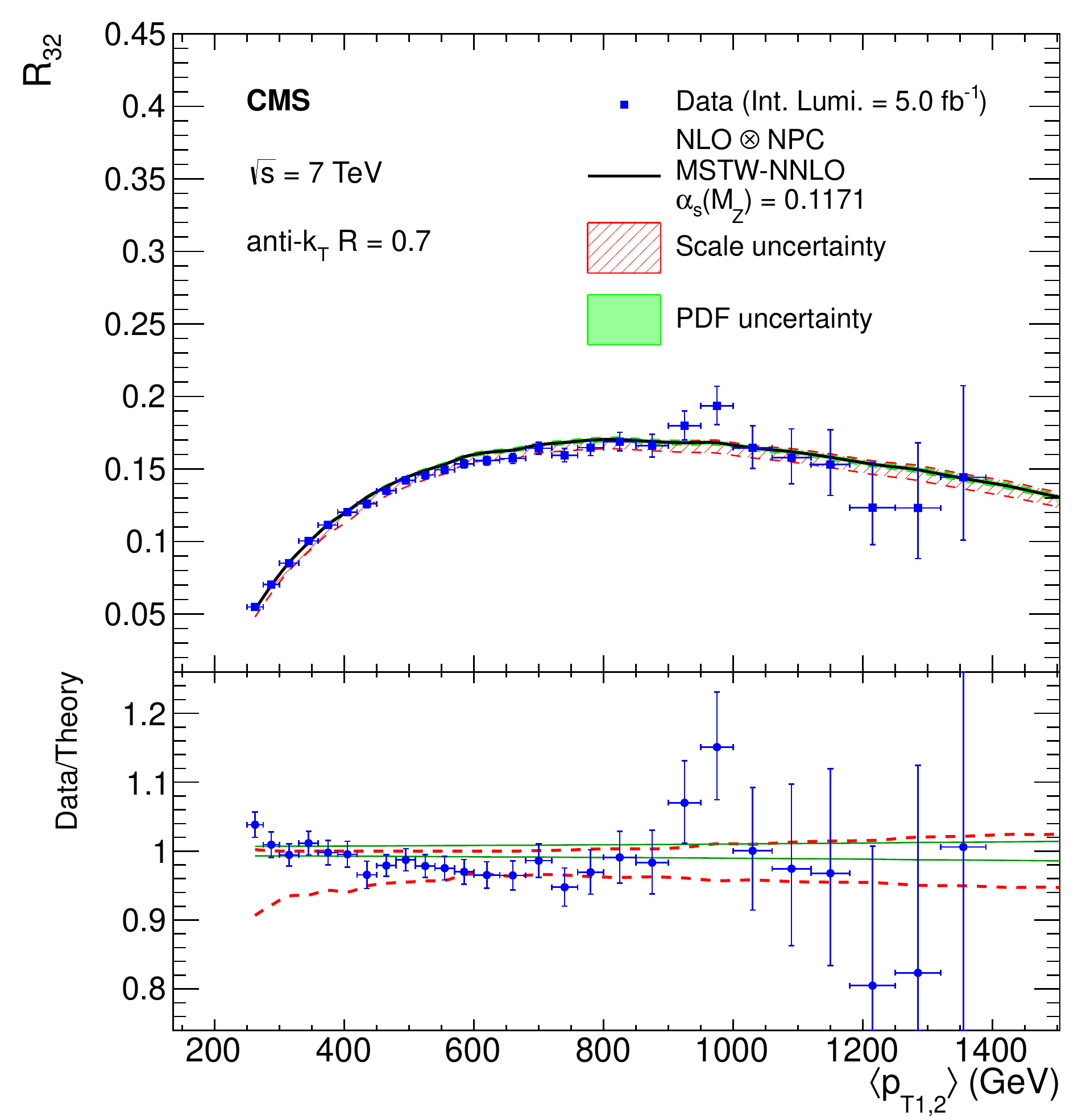}
        \includegraphics[width=0.47\linewidth]{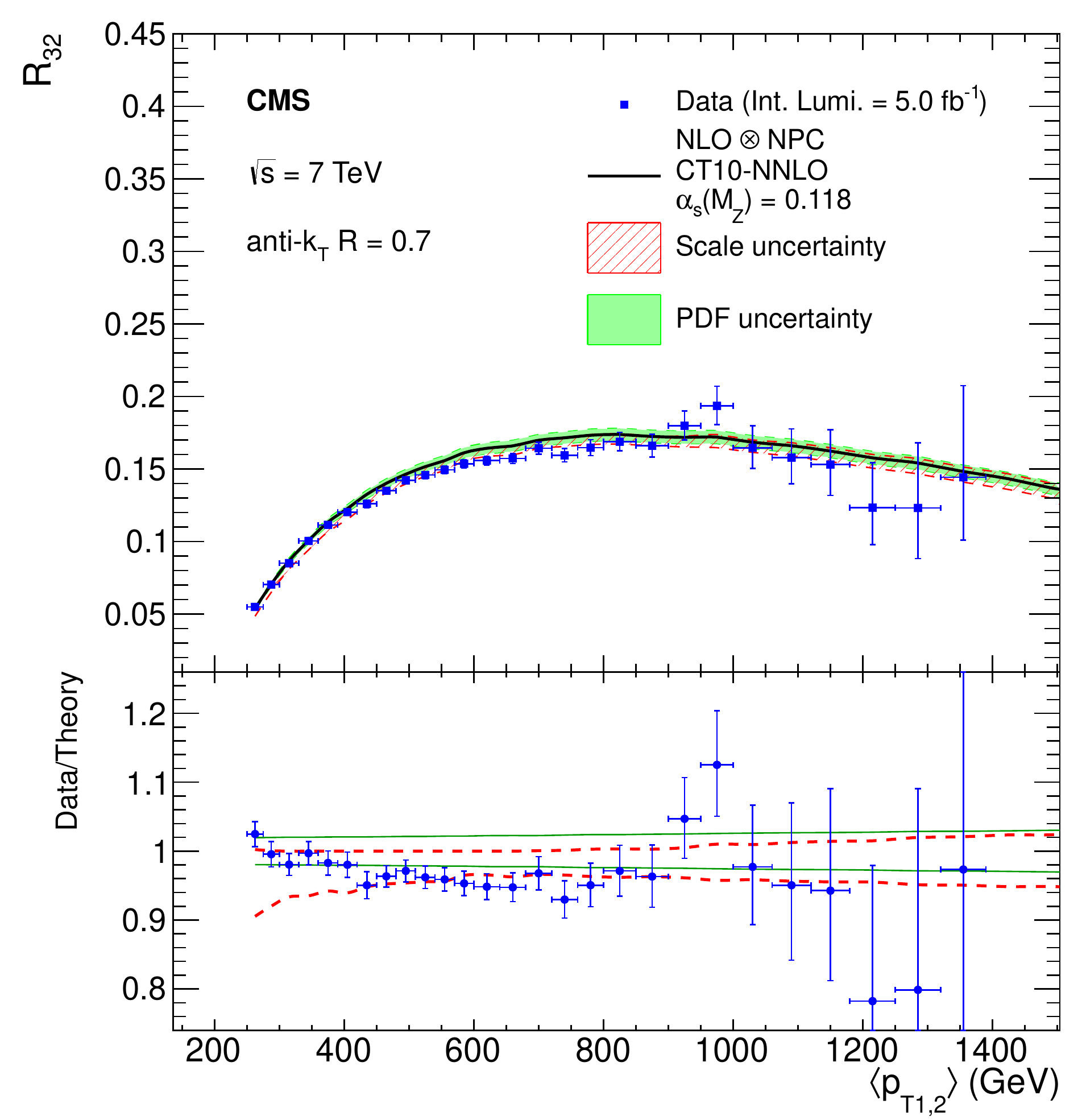}
    \caption{Measurement of \Ratio and NLO predictions using the NNPDF2.1 (top left),
      the ABM11 (top right), the MSTW2008 (bottom left), and the CT10
      (bottom right) NNLO PDF sets. In the upper panel of each plot, the ratio
      \Ratio (solid circles) together with the NLO prediction (solid line) corrected for
      nonperturbative effects (NPC), the scale uncertainty, and the PDF uncertainty
      are shown. The bottom panels show the ratio of data
      to the theoretical predictions, together with bands
      representing the scale (dotted lines) and PDF (solid
      lines) uncertainties. The error bars correspond to the total
      uncertainty. For each PDF set the respective default value of
      \alpsmz is used as indicated.}
    \label{fig:DataTheoryPDFsets_NNPDF_ABM11_MSTW_CT10}
  \end{center}
\end{figure*}

Figure~\ref{fig:DataTheoryPDFsets_NNPDF_ABM11_MSTW_CT10} presents the
measured ratio \Ratio together with NLO predictions using the NNPDF2.1
(top left), the ABM11 (top right), the MSTW2008 (bottom left), and the
CT10 (bottom right) NNLO PDF sets.
The upper panel of each plot shows the ratio
\Ratio (solid circles) together with the NLO prediction (solid line)
corrected for nonperturbative effects,
the scale uncertainty, and the PDF uncertainty.
At the bottom of each plot, the ratio of data over theory
is shown together with bands representing the scale (dotted lines)
and PDF uncertainties (solid lines). The error bars in the figure
correspond to the total uncertainty, for which the statistical and
systematic uncertainties are added in quadrature. For each PDF
set the respective default value of \alpsmz is used in this
comparison as indicated.

The measured ratio rises with increasing \avept as the phase space
opens up for the production of a third jet, reaching a plateau value
for $600 < \avept < 1000\GeV$. At higher \avept \Ratio decreases again
because of the running of \alps, smaller gluon fractions in the total parton luminosity, and
because 3-jet configurations reach kinematic limits earlier than dijet
events.

Scale uncertainties have a very similar behaviour for all PDF sets and
dominate the region up to $\avept \approx 400\GeV$. A comparison to
jets with a size parameter of $R=0.5$
reveals consistent results but with larger scale
uncertainties.

The PDF uncertainties are different for each individual PDF set. The CT10 set
exhibits the largest PDF uncertainties, which are of the order of 2\%
at $\avept = 400\GeV$, increasing to 2.5\% in the 1\TeV region. For
the NNPDF2.1 set PDF uncertainties are of the order of 1.5\% at 400\GeV,
increasing to 2.3\% at 1\TeV. Finally, for the MSTW2008 and ABM11 sets
PDF uncertainties are of the order of 1\% throughout the range of this
measurement.

The comparison of data with the predictions of pQCD in
Fig.~\ref{fig:DataTheoryPDFsets_NNPDF_ABM11_MSTW_CT10} demonstrates
that the NLO calculations using the NNPDF2.1, MSTW2008, and CT10 PDF
sets are in agreement with the measured ratio \Ratio throughout the
range of this measurement. The NLO result employing the ABM11 PDF set
underestimates \Ratio, especially for $\avept < 600\GeV$.

\section{Determination of
  \texorpdfstring{\alpsmz}{alpha-S(M(Z))}\label{section:as}}

\begin{figure*}[hbtp]
  \begin{center}
        \includegraphics[width=0.47\linewidth]{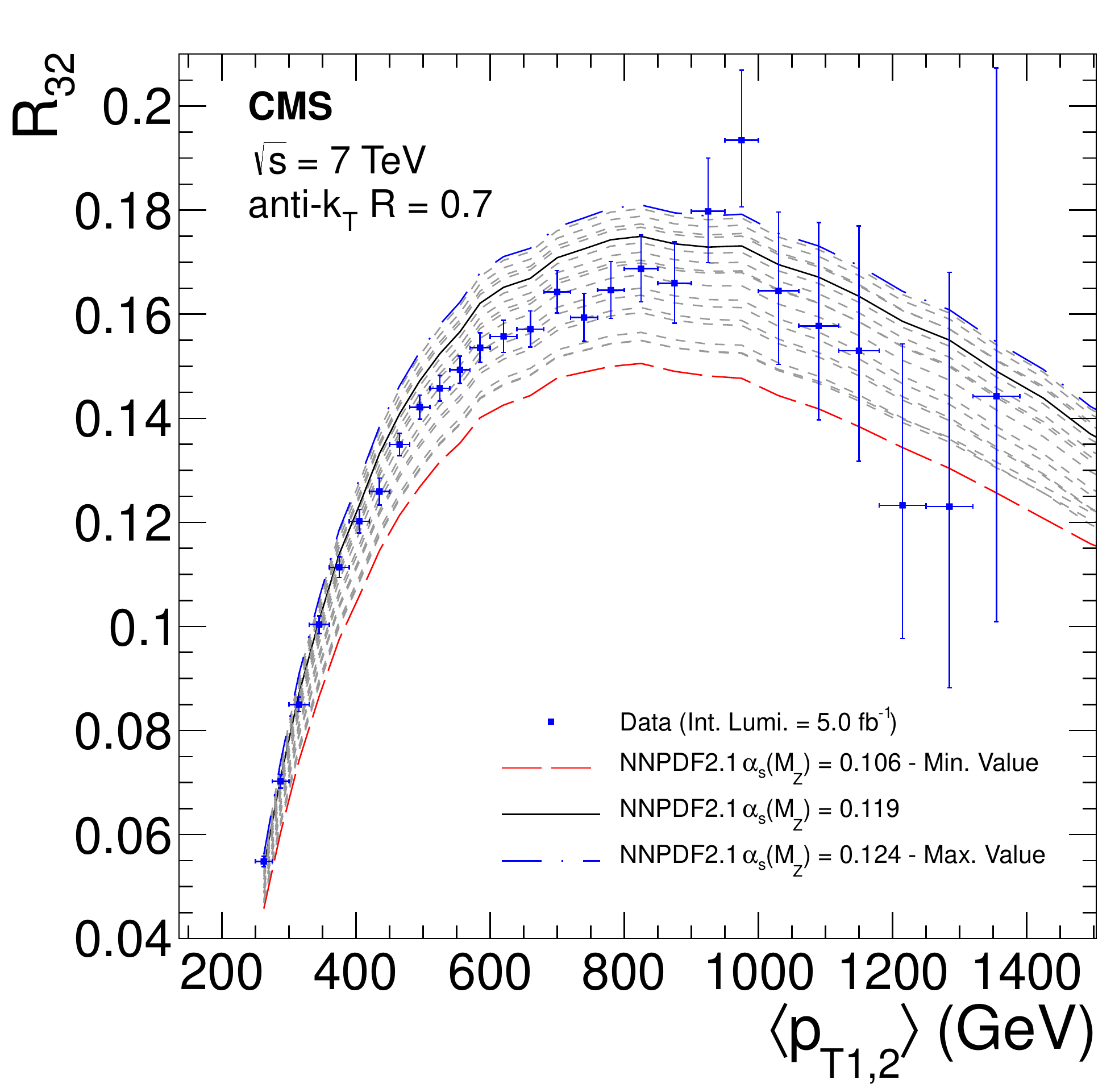}
         \includegraphics[width=0.47\linewidth]{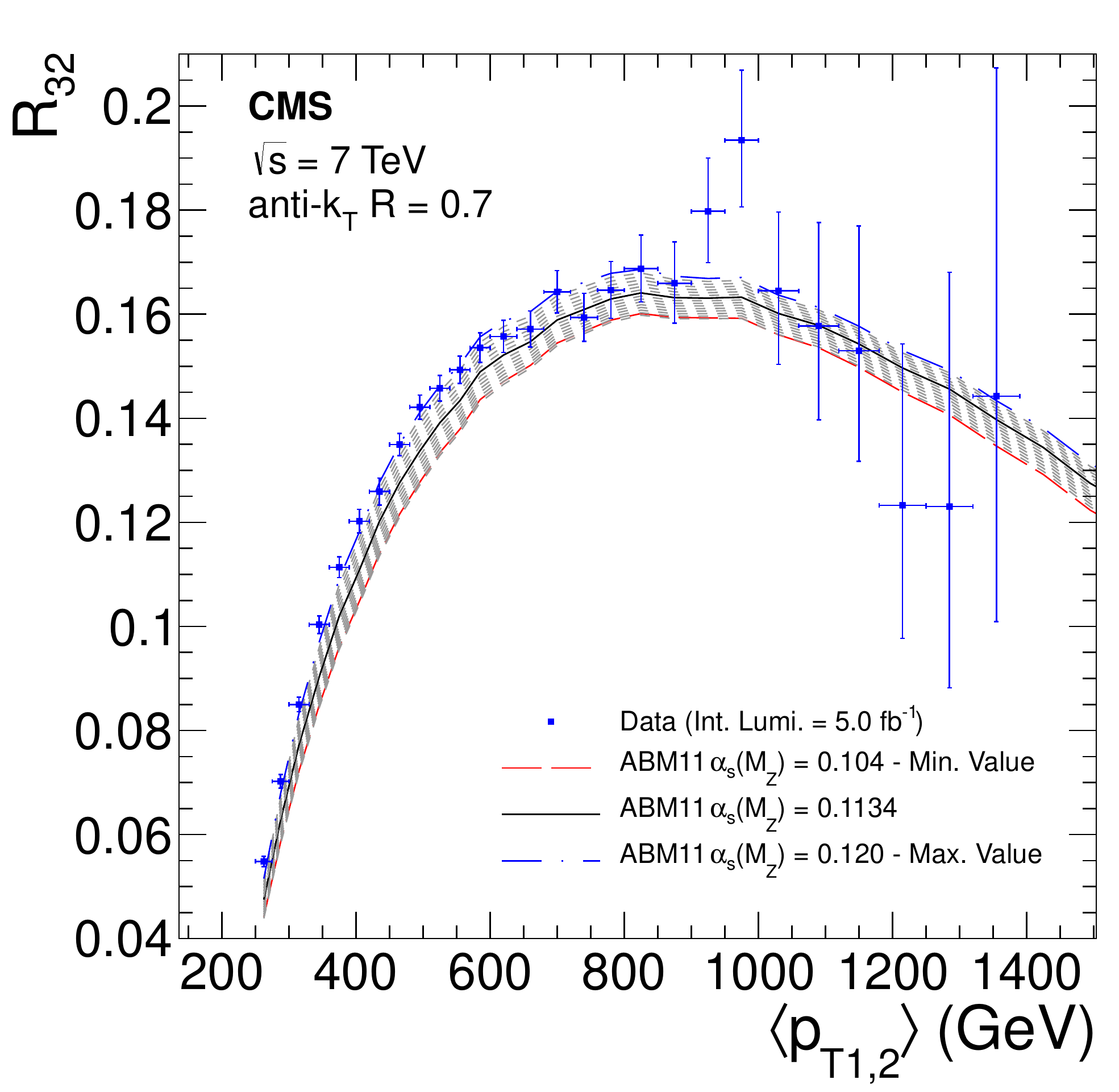}
        \includegraphics[width=0.47\linewidth]{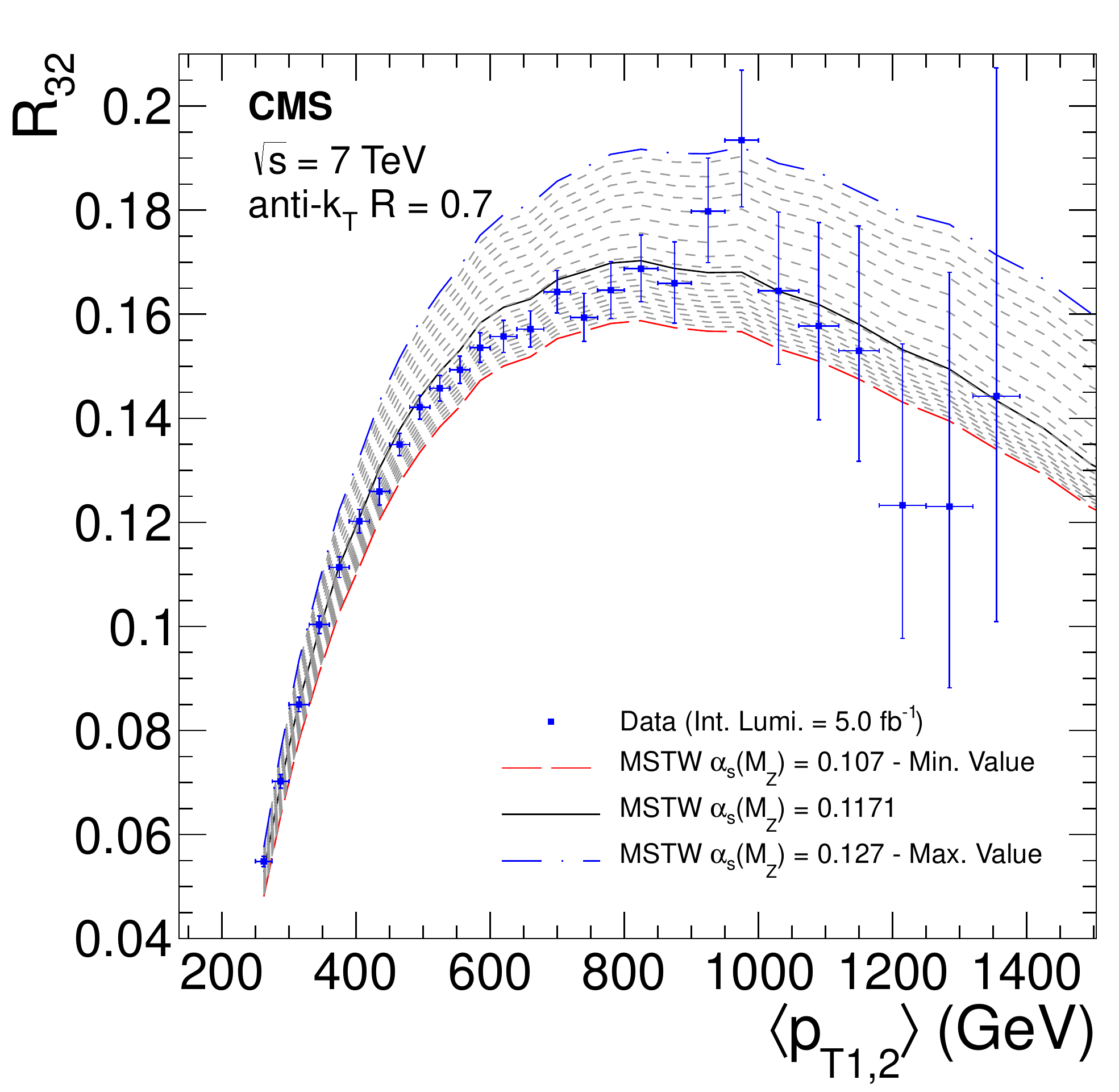}
        \includegraphics[width=0.47\linewidth]{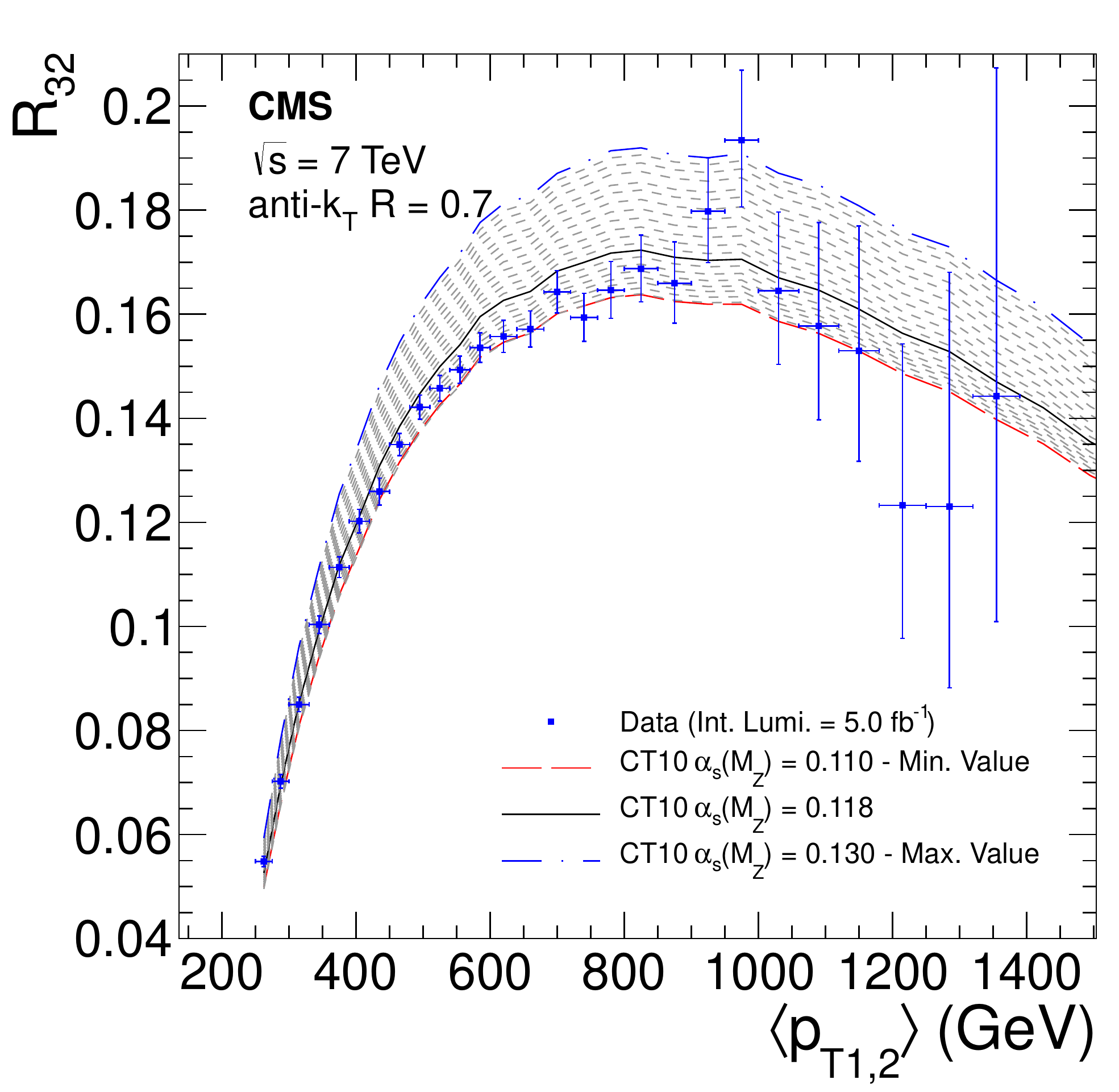}
    \caption{The NLO predictions using the NNPDF2.1 (top left), the
      ABM11 (top right), the MSTW2008 (bottom left), and the CT10
      (bottom right) NNLO PDF sets for a series of values of \alpsmz,
      together with the measured \Ratio. The \alpsmz value is
      varied in the range 0.106--0.124, 0.104--120, 0.107--0.127, and
      0.110--0.130 in steps of 0.001 for the NNPDF2.1, ABM11, MSTW2008, and
      CT10 PDF sets, respectively.}
    \label{fig:as_PDF_sets_NNPDF_ABM11_MSTW_CT10}
  \end{center}
\end{figure*}

The measurement of the ratio \Ratio is used for the determination of
the strong coupling constant \alpsmz.
Figure~\ref{fig:as_PDF_sets_NNPDF_ABM11_MSTW_CT10} shows the
predictions using the NNPDF2.1 (top left), the ABM11 (top right), the
MSTW2008 (bottom left), and the CT10 (bottom right) NNLO PDF sets for
a series of values of \alpsmz, together with the measured \Ratio. The
\alpsmz value is varied in the range 0.106--0.124, 0.104--0.120,
0.107--0.127, and 0.110--0.130 in steps of 0.001 for the NNPDF2.1, ABM11,
MSTW2008, and CT10 PDF sets, respectively.

From Fig.~\ref{fig:as_PDF_sets_NNPDF_ABM11_MSTW_CT10} one observes
that the sensitivity of the ratio \Ratio to variations of the strong
coupling by $\Delta\alpsmz = \pm 0.001$ is different for each of the
four PDF sets. This translates into differences in the
experimental uncertainty in the value of \alpsmz obtained for each PDF
set.

The value of \alpsmz is determined by minimizing the $\chi^{2}$
between the experimental measurement and the theoretical
predictions. The $\chi^{2}$ is defined as
\begin{equation}
  \label{eqn:chi_square}
  \chi^{2}=M^{T}C^{-1}M,
\end{equation}
where $M$ is the vector of the differences between the data
($R_{32}^{i}$) and the theoretical values ($T_{32}^{i}$) in each bin
$i$,
\begin{equation}
  \label{eqn:M_matrix}
  M^{i}=R_{32}^{i}-T_{32}^{i},
\end{equation}
and $C$ is the covariance matrix including all experimental (statistical,
JES, and unfolding) uncertainties. $C$ is defined as
\begin{equation}
  \label{eqn:C_matrix}
  C = \mathrm{Cov}^\text{Stat}+\sum\mathrm{Cov}^\text{JES Sources} +
  \sum\mathrm{Cov}^\text{Unfolding Sources},
\end{equation}
where $\mathrm{Cov}^\mathrm{Stat}$ is the statistical covariance
matrix that accounts for the correlations due to unfolding, and
$\mathrm{Cov}^\text{JES Sources}$,
$\mathrm{Cov}^\text{Unfolding Sources}$ are the covariance matrices
that account for the JES and unfolding systematic uncertainty
sources, respectively. Each systematic uncertainty source for the JES
and unfolding is treated as 100\% correlated across the \avept bins.

To avoid the region with large scale uncertainties close to the
minimal jet \pt requirements, visible in
Fig.~\ref{fig:DataTheoryPDFsets_NNPDF_ABM11_MSTW_CT10}, \alpsmz is
extracted only for $\avept > 420\GeV$. The central result is obtained
by minimizing the $\chi^{2}$ (Eq.~(\ref{eqn:chi_square})) with respect
to \alpsmz for the NNPDF2.1 PDF set, which is the only one that permits
the propagation of PDF uncertainties to the fits for each value of \alpsmz.
The experimental uncertainties are obtained from the \alpsmz values
for which $\chi^{2}$ is increased by one with respect to the minimum
value. The result of a fit to the region of 420--1390\GeV is

\begin{equation}
  \label{eqn:result_exp_NNPDF}
  \begin{split}
    \alpsmz = 0.1148 \pm 0.0014 \,\text{(exp.)} ,
  \end{split}
\end{equation}

with $\chi^{2}/N_\text{dof} = 22.0/20$ at minimum. The experimental
uncertainty contains the statistical, JES, and unfolding sources
(Eq.~(\ref{eqn:C_matrix})), with the JES uncertainty being the
dominant one.

The contribution of PDFs to the uncertainty of the measurement is
evaluated by repeating the fit for each of the 100 PDF replicas of the
NNPDF2.1 set at the relevant value for \alpsmz. In this way 100
determinations of \alpsmz are obtained, whose distribution corresponds
to the propagation of the underlying probability density from the PDFs
to the fitted strong coupling. The PDF uncertainty of the measurement
is then computed as the standard deviation of this distribution. A
more detailed description of the method can be found in
Ref.~\cite{bib:PDF_unc}.

The uncertainties due to the renormalization and factorization scales
are treated separately by varying the default choice of
$\mu_{r}=\mu_{f}=\avept$ between $\avept/2$ and $2\avept$ in six
combinations as explained in Section~\ref{section:measurement}. The
$\chi^{2}$ minimization with respect to \alpsmz is repeated for these
six combinations. The contribution from the $\mu_{r}$, $\mu_{f}$ scale
variations to the uncertainty in the measurement is evaluated by
considering the differences between the NNPDF2.1 \alpsmz central value
and the highest and lowest values found in these six scale
combinations. Out of all scale combinations the lowest \alpsmz value
corresponds to the default scale choice of $\mu_{r}=\mu_{f}=\avept$
and the highest to the scale choice of $\mu_{r}=\mu_{f}=\avept/2$.
The frequent observation of asymmetric scale uncertainties with larger
downward uncertainties in the case of NLO cross sections is
transformed into a purely upward uncertainty for the ratio, as can be
seen in Table~\ref{tbl:as_values_scale}.

\begin{table}[htbH]
  \begin{center}
    \topcaption{The values of \alpsmz at the central scale and for the
      six scale factor combinations.\label{tbl:as_values_scale}}
    \begin{tabular}{cccc}
      \hline\hline
      $\mu_{r}/\avept$  & $\mu_{f}/\avept$ & $\alpsmz \pm \,\text{(exp.)}$ & $\chi^{2}/N_\text{dof}$\rbthm\\\hline
      $1$   \rbtrr      & $1$              & $0.1148 \pm 0.0014 $ & $22.0/20$\\
      $1/2$ \rbtrr      & $1/2$            & $0.1198 \pm 0.0021 $ & $30.6/20$\\
      $1/2$ \rbtrr      & $1$              & $0.1149 \pm 0.0014 $ & $22.2/20$\\
      $1$ \rbtrr        & $1/2$            & $0.1149 \pm 0.0014 $ & $22.2/20$\\
      $1$ \rbtrr        & $2$              & $0.1150 \pm 0.0015 $ & $21.9/20$\\
      $2$ \rbtrr        & $1$              & $0.1159 \pm 0.0014 $ & $20.7/20$\\
      $2$ \rbtrr        & $2$              & $0.1172 \pm 0.0018 $ & $21.3/20$\\
      \hline\hline
    \end{tabular}
  \end{center}
\end{table}

A cross check on the impact of the top quark by imposing $N_f=6$
massless flavours in the NLO matrix elements revealed an increase by
$+0.0009$ in the fitted value of \alpsmz. Further effects, for example from
the evolution of \alps and the PDFs with five or six flavours,
multijet production via fully hadronic decays in the reaction $pp
\rightarrow t\bar{t}+X$, or an incomplete cancellation of electroweak
corrections between numerator and denominator, are estimated to
contribute each at a $\pm 1$\% level to the theoretical uncertainty.
These residual effects are taken into account by symmetrizing the
scale uncertainty such that the largest deviation is adopted as the
total symmetric theory uncertainty.

The final result is

\ifthenelse{\boolean{cms@external}}{
\begin{equation}
  \label{eqn:analytic_result}
  \begin{split}
    \alpsmz = 0.1148 \pm 0.0014\,\text{(exp.)}  \pm
    0.0018\,\text{(PDF)}\\\pm 0.0050\,\text{(theory)},
  \end{split}
\end{equation}
}{
\begin{equation}
  \label{eqn:analytic_result}
    \alpsmz = 0.1148 \pm 0.0014\,\text{(exp.)}  \pm
    0.0018\,\text{(PDF)} \pm 0.0050\,\text{(theory)},
\end{equation}
}

in agreement with the world average value of $\alpsmz = 0.1184 \pm
0.0007$~\cite{bib:PDG2012}, with the Tevatron
results~\cite{bib:as_CDF,bib:as_D0,bib:as_D0_angular}, and a recent
result obtained with LHC data~\cite{bib:as_ATLAS_xsection}.

The determination of \alpsmz, which is based on the NNPDF2.1 PDF set,
is also in agreement with the results obtained using the MSTW2008 or
CT10 PDF sets

\begin{equation}
  \label{eqn:result_exp_other_PDF}
  \begin{split}
    \mathrm{MSTW2008{:}}\;\alpsmz =\, & 0.1141 \pm 0.0022\,\text{(exp.)},\\
    \mathrm{CT10{:}}\;\alpsmz =\, & 0.1135 \pm
    0.0019\,\text{(exp.)},
  \end{split}
\end{equation}

with $\chi^{2}/N_\text{dof} = 20.6/20$ and $21.1/20$, respectively.
If PDF sets with NLO evolution are used instead the impact on
the results of the fits to the ratio observable \Ratio is negligible.
This is in contrast to fits to cross sections, where NNLO
PDF sets usually lead to smaller values of \alpsmz than NLO ones, and
confirms the reduced dependence of \Ratio on details of the PDF
evolution.

In the case of the ABM11 PDFs the series in values of \alpsmz
ends at $0.120$, which is insufficient for a derivation
of the complete shape of the $\chi^2$ curve at minimum such that a fit
value for \alpsmz including uncertainties can only be extrapolated to
give around $\alpsmz = 0.1214 \pm 0.0020 \,\text{(exp.)}$ with
$\chi^{2}/N_\text{dof} = 20.6/20$. For the ABM11 PDF set at NLO with
a default value of $\alpsmz = 0.118$, the series in \alpsmz values ends
at $0.130$ such that a fit can be performed which yields $\alpsmz =
0.1214 \pm 0.0018 \,\text{(exp.)}$, consistent with the extrapolation
above. The fit exhibits, however, a somewhat larger value of
$\chi^{2}/N_\text{dof} = 28.5/20$ compared to the other results.

It is observed that with ABM11 PDFs a higher value of \alpsmz is
preferred. This is in accord with the fact that the ABM11 gluon
density in the phase space relevant for this analysis is significantly
smaller than that of all other PDF sets. Thus, the fit favors a larger
\alpsmz value to compensate for this effect. In summary, the ABM11 PDF
set does not describe the data as well as the alternative PDF sets,
as shown in Fig.~\ref{fig:DataTheoryPDFsets_NNPDF_ABM11_MSTW_CT10}, which
leads to an inferior fit quality and a less consistent result for the
strong coupling.

To investigate the running of the strong coupling constant in more detail,
the fitted region of 420--1390\GeV is split into three bins of \avept
and the fitting procedure is repeated in each of these bins. The three
separate extractions of \alpsmz are reported in
Table~\ref{tbl:as_values}.
The experimental uncertainties in the three obtained values are correlated.
These \alpsmz determinations are
then evolved back to the corresponding values $\alpha_{S}(Q)$ using
the 3-loop solution to the RGE from the NNPDF2.1 set.
For each fit region the cross-section-weighted average of \avept
from the inclusive dijet calculation at NLO with \NLOJETPP is chosen as
the momentum scale $Q$ and is computed to be $Q = 474$,
$664$, and $896\GeV$, respectively. These values, derived again with
the \fastNLO framework, are identical within about 1\GeV for different
PDFs and vary at most by a few \GeVns when using inclusive 3-jet events.

To emphasize that theoretical uncertainties limit the achievable
precision, Table~\ref{tbl:as_unc} presents the decomposition of the
total uncertainty for the three bins in \avept into the experimental,
PDF, and theory components.

\begin{figure}[hbtp]
  \begin{center}
    \includegraphics[width=\cmsFigWidth]{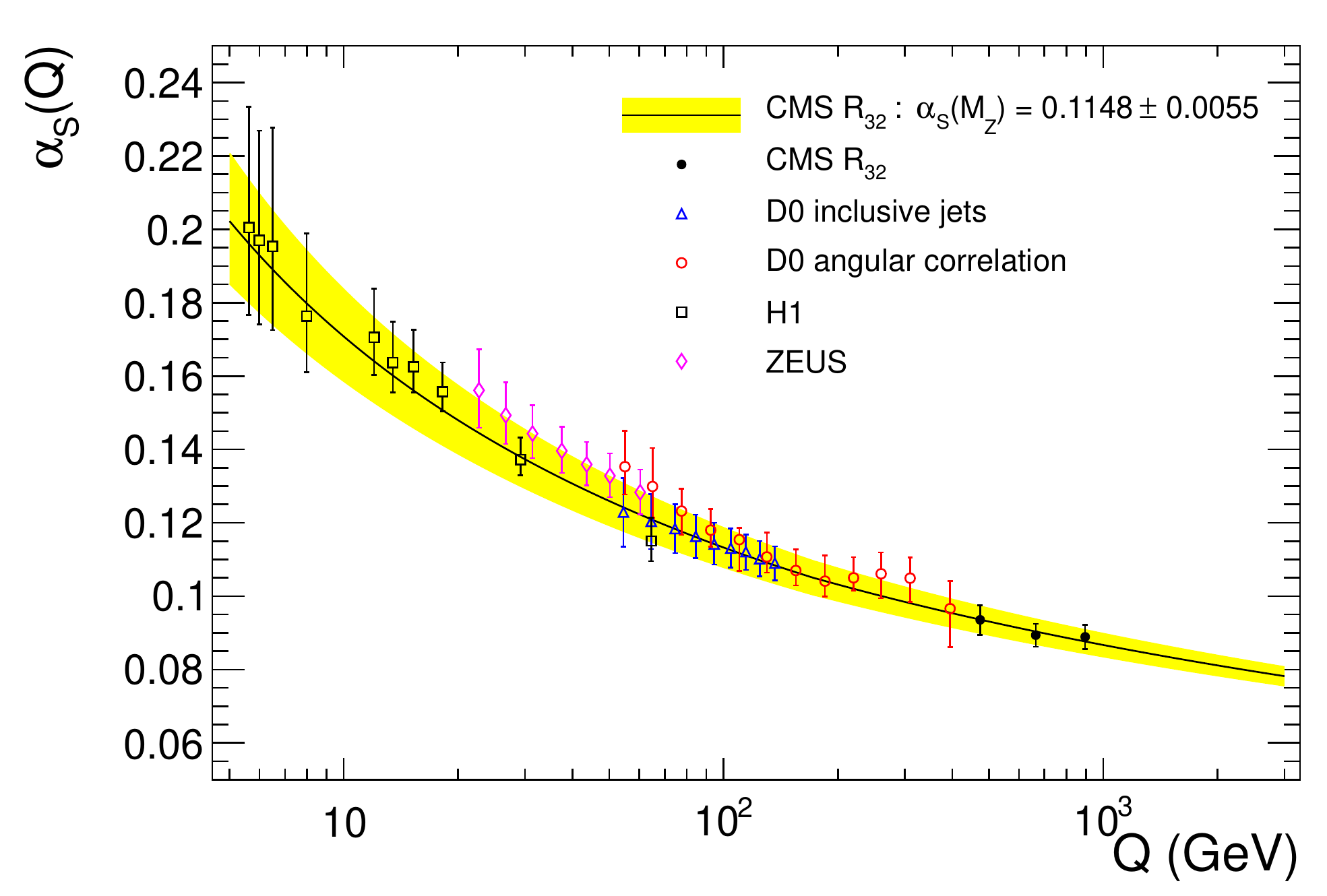}
    \caption{The strong coupling \alpsq (solid line) and its total
      uncertainty (band) evolved from the CMS determination
      $\alpsmz=0.1148 \pm 0.0055$ using a 3-loop solution to
      the RGE as a
      function of the momentum transfer $Q=\avept$. The extractions of
      \alpsq in three separate ranges of $Q$ as presented in
      Table~\ref{tbl:as_values} are shown together with
      results from the H1~\cite{bib:H1as2,bib:H1as1},
      ZEUS~\cite{bib:ZEUSas}, and
      D0~\cite{bib:as_D0,bib:as_D0_angular} experiments at the HERA
      and Tevatron colliders.}
    \label{fig:as_running}
  \end{center}
\end{figure}

Figure~\ref{fig:as_running} presents the strong coupling \alpsq (solid
line) and its total uncertainty (band) as evolved from the CMS
determination, $\alpsmz=0.1148 \pm 0.0055$, using the
3-loop solution to the RGE from NNPDF2.1, as before.
The extractions of \alpsq in three separate ranges of $Q$ as
presented in Table~\ref{tbl:as_values} are also shown.
In the same figure the values of \alps at lower scales determined by
the H1~\cite{bib:H1as2, bib:H1as1}, ZEUS~\cite{bib:ZEUSas}, and
D0~\cite{bib:as_D0, bib:as_D0_angular} Collaborations are shown for
comparison.
The results on \alps reported here are consistent with the
energy dependence predicted by the RGE and extend the range,
in which the RGE is tested, to the region of several hundred \GeV.

\begin{table*}[htbH]
  \begin{center}
    \topcaption{The separate determinations of \alps in bins of
      \avept.\label{tbl:as_values}}
    \begin{tabular}{cccccc}
      \hline\hline
      \avept range   & $Q$      & \alpsmz                        & \alpsq\rbthm                   & No.\ of data & $\chi^{2}/N_\text{dof}$ \\
      (\GeVns{})       & (\GeVns{}) &                                &                                & points      & \\\hline
      420--600 \rbtrr & 474    & $0.1147\pm{0.0061}$ & $0.0936\pm{0.0041}$ &  6\rbtrr    & $4.4/5$\\
      600--800 \rbtrr & 664    & $0.1132\pm{0.0050}$ & $0.0894\pm{0.0031}$            &  5\rbtrr    & $5.9/4$\\
      800--1390\rbtrr & 896    & $0.1170\pm{0.0058}$ & $0.0889\pm{0.0034}$ & 10\rbtrr    & $5.7/9$\\
      \hline\hline
    \end{tabular}
  \end{center}
\end{table*}

\begin{table*}[htbH]
  \begin{center}
    \topcaption{Uncertainty composition for \alpsmz from the
      determination of \alps in bins of \avept.\label{tbl:as_unc}}
    \begin{tabular}{cccccc}
      \hline\hline
      \avept range    & $Q$      & \alpsmz & exp.\rbthm    & PDF           & theory\\
      (\GeVns{})        & (\GeVns{}) &         &               &               & \\\hline
      420--600 \rbtrr & 474    & 0.1147  & $\pm{0.0015}$ & $\pm{0.0015}$ & $\pm{0.0057}$\\
      600--800 \rbtrr & 664    & 0.1132  & $\pm{0.0018}$ & $\pm{0.0025}$ & $\pm{0.0039}$\\
      800--1390\rbtrr & 896    & 0.1170  & $\pm{0.0024}$ & $\pm{0.0021}$ & $\pm{0.0048}$\\
      \hline\hline
    \end{tabular}
  \end{center}
\end{table*}

\section{Summary\label{section:summary}}

The ratio \Ratio of the inclusive 3-jet cross section to the inclusive
2-jet cross section, for jets with $\pt > 150\GeV$ and $\vert y \vert
< 2.5$, has been measured in the range $250 < \avept < 1390\GeV$ for
proton-proton collisions at a centre-of-mass energy of 7\TeV.  The
results have been compared with predictions of QCD at NLO obtained
with various PDF sets. The NLO calculations using the NNPDF2.1,
MSTW2008, and CT10 NNLO PDF sets are in agreement with the measured ratio
\Ratio throughout the range of \avept studied. However, calculations
using the ABM11 PDF sets underestimate \Ratio for $\avept<600\GeV$.

Measurements of \Ratio over the range $420 < \avept < 1390\GeV$ have
been used to determine the strong coupling constant \alps at the scale
of the \cPZ\ boson mass.
The final result is

\ifthenelse{\boolean{cms@external}}{
\begin{equation*}
  \label{eqn:total_result}
  \begin{split}
    \alpsmz = 0.1148 \pm 0.0014\,\text{(exp.)}  \pm
    0.0018\,\text{(PDF)}\\ \pm 0.0050\,\text{(theory)} =
    0.1148 \pm 0.0055,
  \end{split}
\end{equation*}
}
{
\begin{equation*}
  \label{eqn:total_result}
    \alpsmz = 0.1148 \pm 0.0014\,\text{(exp.)}  \pm
    0.0018\,\text{(PDF)} \pm 0.0050\,\text{(theory)} =
    0.1148 \pm 0.0055,
\end{equation*}
}

where experimental, PDF, and theory uncertainties have been added
quadratically to give the total uncertainty. The result is in
agreement with the world average value of
$\alpsmz=0.1184\pm0.0007$~\cite{bib:PDG2012} and represents the first
determination of the strong coupling constant from jet measurements with
momenta of the order of 1\TeV. The dominant uncertainties are of
theoretical origin and limit the currently achievable precision. The
predicted ratio depends only indirectly on the evolution of the parton
distribution functions of the proton and consequently this measurement
also serves as a test of the evolution of \alpsq. No
deviation from the expected behaviour is observed.

\section*{Acknowledgements}

{\tolerance=500
We thank Gavin Salam for his valuable comments to this paper.
We congratulate our colleagues in the CERN accelerator departments for the excellent performance of the LHC and thank the technical and administrative staffs at CERN and at other CMS institutes for their contributions to the success of the CMS effort. In addition, we gratefully acknowledge the computing centres and personnel of the Worldwide LHC Computing Grid for delivering so effectively the computing infrastructure essential to our analyses. Finally, we acknowledge the enduring support for the construction and operation of the LHC and the CMS detector provided by the following funding agencies: BMWF and FWF (Austria); FNRS and FWO (Belgium); CNPq, CAPES, FAPERJ, and FAPESP (Brazil); MEYS (Bulgaria); CERN; CAS, MoST, and NSFC (China); COLCIENCIAS (Colombia); MSES (Croatia); RPF (Cyprus); MoER, SF0690030s09 and ERDF (Estonia); Academy of Finland, MEC, and HIP (Finland); CEA and CNRS/IN2P3 (France); BMBF, DFG, and HGF (Germany); GSRT (Greece); OTKA and NKTH (Hungary); DAE and DST (India); IPM (Iran); SFI (Ireland); INFN (Italy); NRF and WCU (Republic of Korea); LAS (Lithuania); CINVESTAV, CONACYT, SEP, and UASLP-FAI (Mexico); MSI (New Zealand); PAEC (Pakistan); MSHE and NSC (Poland); FCT (Portugal); JINR (Armenia, Belarus, Georgia, Ukraine, Uzbekistan); MON, RosAtom, RAS and RFBR (Russia); MSTD (Serbia); SEIDI and CPAN (Spain); Swiss Funding Agencies (Switzerland); NSC (Taipei); ThEPCenter, IPST and NSTDA (Thailand); TUBITAK and TAEK (Turkey); NASU (Ukraine); STFC (United Kingdom); DOE and NSF (USA).

Individuals have received support from the Marie-Curie programme and the European Research Council and EPLANET (European Union); the Leventis Foundation; the A. P. Sloan Foundation; the Alexander von Humboldt Foundation; the Belgian Federal Science Policy Office; the Fonds pour la Formation \`a la Recherche dans l'Industrie et dans l'Agriculture (FRIA-Belgium); the Agentschap voor Innovatie door Wetenschap en Technologie (IWT-Belgium); the Ministry of Education, Youth and Sports (MEYS) of Czech Republic; the Council of Science and Industrial Research, India; the Compagnia di San Paolo (Torino); the HOMING PLUS programme of Foundation for Polish Science, cofinanced by EU, Regional Development Fund; and the Thalis and Aristeia programmes cofinanced by EU-ESF and the Greek NSRF.
\par}

\bibliography{auto_generated}   

\cleardoublepage \appendix\section{The CMS Collaboration \label{app:collab}}\begin{sloppypar}\hyphenpenalty=5000\widowpenalty=500\clubpenalty=5000\textbf{Yerevan Physics Institute,  Yerevan,  Armenia}\\*[0pt]
S.~Chatrchyan, V.~Khachatryan, A.M.~Sirunyan, A.~Tumasyan
\vskip\cmsinstskip
\textbf{Institut f\"{u}r Hochenergiephysik der OeAW,  Wien,  Austria}\\*[0pt]
W.~Adam, T.~Bergauer, M.~Dragicevic, J.~Er\"{o}, C.~Fabjan\cmsAuthorMark{1}, M.~Friedl, R.~Fr\"{u}hwirth\cmsAuthorMark{1}, V.M.~Ghete, N.~H\"{o}rmann, J.~Hrubec, M.~Jeitler\cmsAuthorMark{1}, W.~Kiesenhofer, V.~Kn\"{u}nz, M.~Krammer\cmsAuthorMark{1}, I.~Kr\"{a}tschmer, D.~Liko, I.~Mikulec, D.~Rabady\cmsAuthorMark{2}, B.~Rahbaran, C.~Rohringer, H.~Rohringer, R.~Sch\"{o}fbeck, J.~Strauss, A.~Taurok, W.~Treberer-Treberspurg, W.~Waltenberger, C.-E.~Wulz\cmsAuthorMark{1}
% ...
\vskip\cmsinstskip
% ...
1:~~Also at Vienna University of Technology, Vienna, Austria\\
2:~~Also at CERN, European Organization for Nuclear Research, Geneva, Switzerland\\
% ...

\end{sloppypar}
\end{document}